\documentclass[%
 reprint,
superscriptaddress,
 amsmath,amssymb,
 aps,
floatfix,
]{revtex4-2}

\usepackage{graphicx}
\usepackage{dcolumn}
\usepackage{bm}
\usepackage{amsmath,amssymb}
\usepackage{textcomp,mathcomp}

\begin{document}

\preprint{APS/123-QED}

\title{Zigzag chain order of LiVSe$_2$ developing away from the vanadium trimer phase transition boundary}



\author{K. Kojima}
\affiliation{Department of Applied Physics, Nagoya University, Aichi 464-8603, Japan}
\author{N. Katayama}\thanks{Corresponding author.}\email{katayama.naoyuki.m5@f.mail.nagoya-u.ac.jp}
\affiliation{Department of Applied Physics, Nagoya University, Aichi 464-8603, Japan}
%
%
\author{K. Sugimoto}
\affiliation{Department of Physics, Keio University, Kanagawa 223-8522, Japan}
%
%
\author{N. Hirao}
\affiliation{Diffraction and Scattering Division, Center for Synchrotron Radiation, Japan Synchrotron Radiation Research Institute, Hyogo 679-5198, Japan}
\author{Y. Ohta}
\affiliation{Department of Physics, Chiba University, Chiba 263-8522, Japan}
\author{H. Sawa}					
\affiliation{Department of Applied Physics, Nagoya University, Aichi 464-8603, Japan}
\date{\today}

\begin{abstract}
The phenomenon of self-assembly of constituent elements to form molecules at low temperatures appears ubiquitously in transition metal compounds with orbital degrees of freedom. Recent progress in local structure studies using synchrotron radiation x-rays is shifting the interest in structural studies in such molecule-forming systems from the low-temperature ordered phase to the short-range order that appears like a precursor at high temperatures. In this study, we discuss both experimentally and theoretically the relationship between the trimer structure that appears in the layered LiV$X_2$ ($X$ = O, S, Se) system with a two-dimensional triangular lattice of vanadium and the zigzag chain-like local structure that appears near the phase transition boundary where molecular formation occurs. The vanadium trimerization that persistently appears in both low-temperature phases of LiVO$_2$ and LiVS$_2$ disappears in LiVSe$_2$, and a regular triangular lattice is thought to be realized in LiVSe$_2$, but this study reveals that the zigzag chain local distortion appears with a finite correlation length. This zigzag chain state local distortions are similar to the motif of local distortions in the high-temperature phase of LiVS$_2$, indicating that the local distortions are persistent away from the trimer phase transition boundary. On the other hand, it is concluded that the zigzag chain order appearing in LiVSe$_2$ is more stable than that in LiVS$_2$ in terms of the temperature variation of atomic displacement and correlation length. The zigzag chain order is considered to be competitive with the trimer order appearing in the LiV$X_2$ system. In this paper, we discuss the similarities and differences between the parameters that stabilize these electronic phases and the local distortions that appear in other molecular formation systems.

\end{abstract}

\maketitle

\section{Introduction}

Since the discovery of the Verwey transition of magnetite in the 1930s \cite{magnetite1}, many physical and structural studies have been devoted to transition metal compounds that undergo molecular formation at low temperatures \cite{orbitally, Khomskii_Sergey_review, Whangbo, Rovira_Whangbo, CuIr2S4, AlV2O4, AlV2O4-2, LiVS$_2$-1, LiVO2-Tian, MgTi2O4, LiRh2O4, LiRh2O4-2, LiMoO2, CsW2O6, Li2RuO3, kobayashi, Li033VS2, RuP1, NaTiSi2O6_1, BaV10O15}. The molecular formation phenomena in transition metal compounds often appear in pyrochlore and triangular lattice compounds with high symmetry. High $d$-orbital degeneracy and associated orbital degrees of freedom, the low dimensionality of electronic structures such as hidden one-dimensionality produced by orbital linkages, and competition between itinerancy and localization are considered to be important factors in molecular formation phenomena \cite{orbitally, Khomskii_Sergey_review, Whangbo, Rovira_Whangbo}, and together with drastic physical properties such as metal-insulator transition and giant entropy change that appear with molecular formation \cite{Li033VS2, VO2_electric, VO2_thermal, LiVO2_PCM}, they have attracted much attention from both basic and applied perspectives. With recent progress in structural analysis techniques using synchrotron radiation x-rays and neutrons, the interest in structural studies of these molecular formation systems is shifting from elucidating the complex molecular formation patterns that appear in the low-temperature phase to the precursor local distortions that appear in the high-temperature phase. For examples, in the spinel lattice system AlV$_2$O$_4$, vanadium heptamer (trimer + tetramer), which appears at low temperatures, persists as short-range order at high temperatures far beyond the phase transition \cite{AlV2O4-2}, and in CuIr$_2$S$_4$, which forms complex dimer patterns with charge ordering at low temperatures, short-ranged tetragonal distortion appears in the high-temperature phase \cite{CuIr2S4_2}, as revealed by a pair distribution function analysis (PDF) using synchrotron radiation x-rays. The existence of such local distortions indicates that orbital ordering is localized in the high-temperature phase, which was believed to maintain high orbital degeneracy \cite{CuIr2S4_2}, and may affect our understanding of the mechanism of molecular formation and the thermodynamics associated with phase transitions. 

\begin{figure}
\includegraphics[width=86mm]{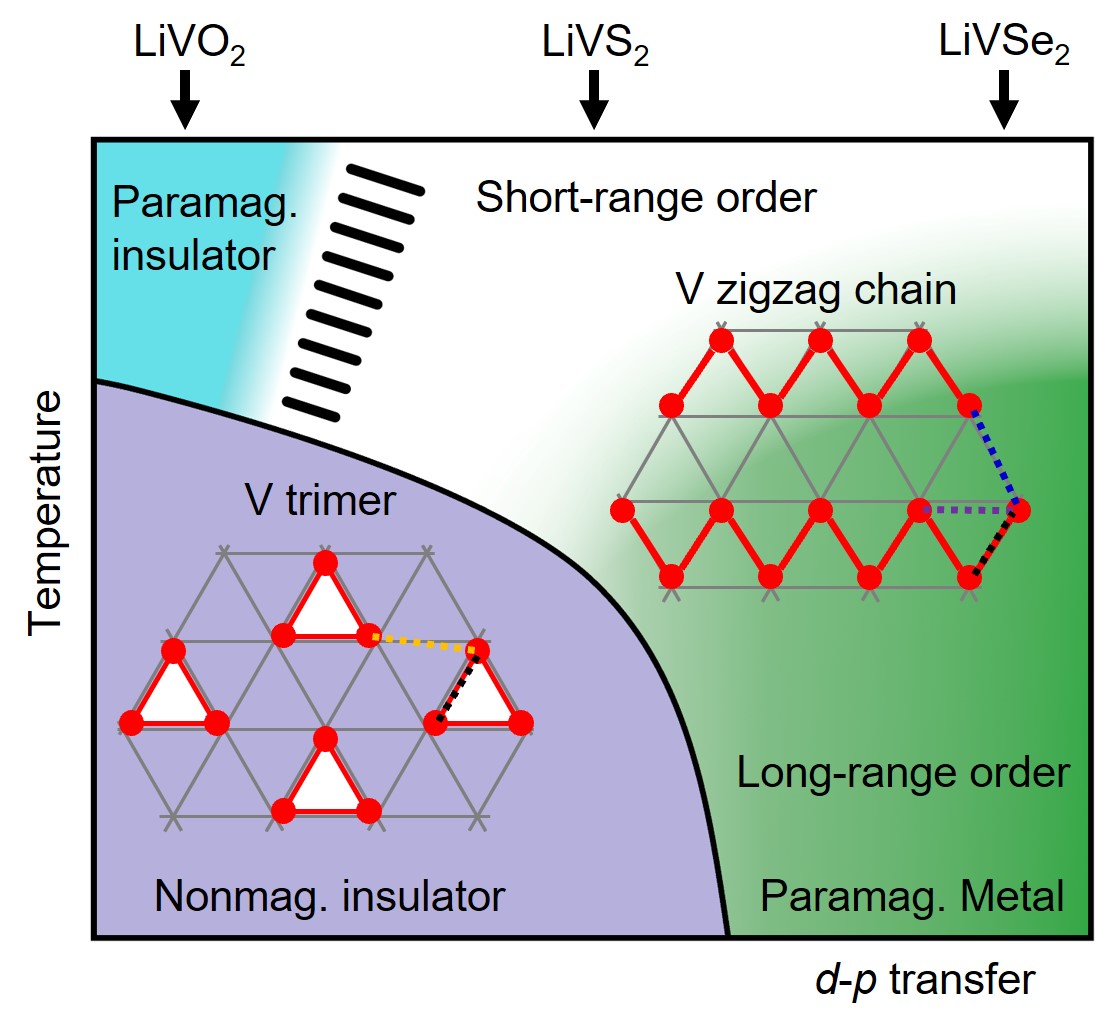}
\caption{\label{fig:Phase_diag} Phase diagram of the LiV$X_2$ system. Although the classification of each phase has been shown in previous studies \cite{LiVS$_2$-1, LiVS$_2$-2}, our study reveals that the zigzag chain state shown in green appears more stable in LiVSe$_2$, which is farther from the trimer phase boundary compared to LiVS$_2$. The different colored dotted lines on the inset figures indicate different V-V distances.}
\end{figure}

Layered LiV$X_2$ ($X$ = O, S, Se) with a two-dimensional triangular lattice provides a unique playground for studying the relationship between the molecular formations that appear in the low-temperature phase and the local distortions that appear in the high-temperature phase. The electronic phase diagram characterizing the LiV$X_2$ system is shown in Fig.~\ref{fig:Phase_diag}, which was modified from Fig.1 in the Ref. \cite{LiVS$_2$-1} based on the zigzag chain order of LiVSe$_2$, which will be presented in this article later. LiVO$_2$ is an insulator in the whole temperature range and shows a nonmagnetic insulator transition at about 480 K. It has long been argued that a vanadium trimer state appears in the low-temperature nonmagnetic insulating state of LiVO$_2$, which has been fully clarified by a recent PDF structural analysis based on the unique structural model \cite{LiVO2-kj, LiVO2-kj2}. Substitution of O(2$p$) for S(3$p$) or Se(4$p$) increases the itinerancy. In LiVS$_2$, the high-temperature phase becomes metallic while maintaining the trimer in the low-temperature phase, and in LiVSe$_2$, the trimer state is completely suppressed, and the metallic state appears at all temperatures.

Similar to AlV$_2$O$_4$ \cite{AlV2O4-2} and CuIr$_2$S$_4$ \cite{CuIr2S4_2} mentioned above, LiVS$_2$ also shows local distortion on the triangular lattice of vanadium above the trimer transition temperature of $T_c$ $\sim$ 314 K \cite{LiVS$_2$-2}. However, the characteristics of the local structure of LiVS$_2$ are very different from those of other materials that undergo molecular formation. First, in the high-temperature phase of LiVS$_2$, a zigzag chain order appears preliminarily, which is clearly different from the trimer motif in the low-temperature phase. Second, the correlation length of the local distortion appearing in the conventional molecular formation system is only a few Å, whereas the zigzag chain order appearing in the high-temperature phase of LiVS$_2$ has a long correlation length of several hundred Å. Surprisingly, the correlation length increases further just above the trimer transition and superlattice peaks appear, which originate from the zigzag chain ordering. Third, this zigzag chain distortion appears to exhibit spatially and temporally fluctuating dynamics when observed by time-resolved Scanning Transmission Electron Microscope (STEM) measurements.

It has been argued that this anomalous local distortion is related to the appearance of pseudogaps, a phenomenon of anomalous reduction in magnetic susceptibility that appears in the metallic phase near the trimer phase transition boundary in terms of physical properties \cite{LiVS$_2$-1, LiVS$_2$-2}. If the zigzag chain order originates from the V$^{3+}$($d^2$) electronic state and implies bond formation between neighboring vanadium ions, it is understood that the evolution of correlation length implies a gradual increase in the number of $d$ electrons involved in spin-singlet formation, leading to a pseudogap-like decrease in magnetic susceptibility. Based on this idea, LiVSe$_2$ could be an attractive research target to investigate the development of zigzag chain order in LiV$X_2$ ($X$ = O, S, Se) phase space. This is because LiVSe$_2$ exhibits metallic conduction over the entire temperature range, and this metallic phase is connected to the metallic phase of LiVS$_2$ in the phase space where the zigzag chains appear \cite{LiVS$_2$-1}. More interestingly, according to the temperature dependence of the magnetic susceptibility of LiVSe$_2$ reported previously \cite{LiVSe$_2$_murphy}, the Pauli paramagnetic component of the magnetic susceptibility obtained by subtracting the Curie tail has a positive slope with respect to temperature, seems to indicate that a pseudogap similar to that of LiVS$_2$ appears persistently in LiVSe$_2$. It should be noted that a similar temperature dependence of magnetic susceptibility has been observed for our LiVSe$_2$ samples, the details of which are summarized in Supplemental Material \cite{SI}. By clarifying the appearance of zigzag chains in LiVSe$_2$, examining their stability with respect to temperature and pressure, and comparing them with the characteristics of zigzag chains in LiVS$_2$, it is expected that the relationship between the trimer phase and zigzag chain order can be studied in detail.

In this paper, we discuss the relationship between the zigzag chain ordering and the trimer phase in the LiV$X_2$ ($X$ = O, S, Se) system, focusing on the results of structural studies obtained utilizing synchrotron radiation x-rays and theoretical calculations of LiVSe$_2$. Synchrotron x-ray diffraction experiments have revealed that the zigzag chain order appears in LiVSe$_2$ with a finite correlation length, as in the high-temperature phase of LiVS$_2$. Strangely, even though LiVSe$_2$ is farther from the trimer transition boundary in phase space, the zigzag chain order has a larger vanadium displacement than that of LiVS$_2$ and persists at higher temperatures with longer correlation lengths. This indicates that the zigzag chain local distortion that appears in the LiV$X_2$ system is not merely a precursor to the trimerization that appears at low temperatures. Based on theoretical calculations and pressure effects, we discuss the factors that stabilize the trimer and zigzag chains, and argue that the LiV$X_2$ system is the stage for studying a unique local distortion phenomenon that is clearly different from conventional molecular formation systems.

\section{Methods}

\subsection{Sample synthesis}

LiVSe$_2$ was synthesized by a three-step synthesis method. First, the raw materials, V (99.5\%) and Se (99.999\%), were mixed so that V:Se = 1:1.01 to avoid V self-intercalation, sealed in a vacuum quartz tube, and calcined at 700 $\tccentigrade$ for 3 days to obtain 1$T$-VSe$_2$. The resulting powder sample was then immersed in a large excess of 0.2 $M$ $n$-BuLi/hexane solution for 2 days in a glove box under an Ar atmosphere. This yields Li$_2$VSe$_2$. To remove excess Li, the sample was immersed in I$_2$-acetonitrile; by adjusting the weight of I$_2$, the deintercalation of Li can be quantitatively controlled. The resulting samples were annealed at 150 $\tccentigrade$ for 10 hours to restore crystallinity and obtain powder samples of LiVSe$_2$. For LiVO$_2$ and LiVS$_2$, we used the same batch of samples as presented in the previous paper \cite{LiVO2-kj}.

\subsection{Experiments}

Inductively coupled plasma (ICP) measurements were performed to calculate the Li content of the synthesized samples. Considering the presence of impurity Li$_2$Se (3.4\%) revealed by Rietveld analysis, ICP measurement estimated the LiVSe$_2$ used in this study to be Li/V = 1.01. The occupancy of the Se site was 0.983(3) as a result of our Rietveld analysis. Differential scanning calorimetry (DSC) was conducted using a DSC 204 F1 Phoenix (Netzsch).

The ambient pressure x-ray diffraction data were measured at BL44B2 and BL02B2 at SPring-8. 30 keV x-rays and the detectors commonly used at each beamline were used. The data obtained at BL44B2 were corrected, combined, and converted to one-dimensional data \cite{BL44B2-1, BL44B2-2}. The x-ray diffraction experiments under high pressure were performed at BL10XU at SPring-8 \cite{BL10XU}. A diamond anvil cell (DAC) was used for high-pressure generation. The sample was loaded into the DAC, together with some rubies as pressure markers \cite{ruby_mao}. Helium was used as a pressure-transmitting medium. 30 keV x-ray energy was used, and diffraction data were acquired with an imaging plate detector. 
The obtained diffraction data were indexed using Conograph \cite{conograph}, and the analysis was performed using Rietan-FP \cite{RIETAN}. The obtained structures were drawn using VESTA \cite{vesta}. 

X-ray absorption fine structure (XAFS) measurements were performed on the vanadium $K$-edge ($\sim$ 5.47 keV) at BL5S2 and BL11S2 at Aichi Synchrotron Radiation Center. XAFS spectra were collected by the Quick-XAFS method using a vanadium foil as a reference. The samples were mixed with BN and pelletized. Temperatures below 300 K were controlled by a cryostat, while temperatures above 300 K were controlled by a heater. The obtained data were transformed using Athena included in Demeter \cite{athena}. For the Fourier transform, the range up to 12 $\mathrm{\AA}^{-1}$ was used for the almost data, and for the 30 K data of LiVO$_2$, the range up to 13 $\mathrm{\AA}^{-1}$ was used so that the interatomic distances originating from the trimer would be apparent.

\subsection{Theoretical calculations}

The partial densities of states (PDOS) and the Fermi surfaces in LiVSe$_2$ and LiVS$_2$ were calculated by employing the WIEN2k code \cite{Blaha2020,WIEN2k} based on the full-potential linearized augmented plane-wave method. These results are obtained using the generalized gradient approximation (GGA) for electron correlations, where the exchange-correlation potential of reference \cite{Perdew1996} is used. To improve the description of electron correlations in V 3$d$ orbitals, we use the GGA+$U$ method with the around-the-mean-field double-counting scheme \cite{Czyzyk1994PRB}, which is believed to be suitable for metallic systems. In the following, we will present the results obtained at $U$ = 4 eV. PDOS for the triangular lattice of LiVS$_2$ and LiVSe$_2$ were calculated using lattice parameters obtained by analyzing LiVS$_2$ data at 400 K and LiVSe$_2$ at 300 K assuming trigonal space group $P\bar{3}m1$. We also calculated the PDOS for the zigzag chain structure of both materials using the lattice parameters obtained by analyzing the data of LiVS$_2$ at 320 K and LiVSe$_2$ at 100 K assuming the original space group, the monoclinic space group $Pm$. It should be noted that monoclinic distortions of these materials were small at high temperatures and could be indexed by the trigonal space groups, as discussed below. In the self-consistent calculations, we use $12 \times 12 \times 6$ $\bm{k}$-points in the Brillouin zone. The muffin-tin radii ($R_{\mathrm{MT}}$) for the $P\bar{3}m1$ structure are taken as 2.23 (Li), 2.45 (V), and 2.11 (S) Bohr for LiVS$_2$ and 2.14 (Li), 2.4 (V), and 2.4 (Se) Bohr for LiVSe$_2$. Similarly, the $R_{\mathrm{MT}}$ for the $Pm$ structure are taken as 2.16 (Li), 2.29 (V), and 2.07 (S) Bohr for LiVS$_2$ and 2.11 (Li), 2.3 (V), and 2.3 (Se) Bohr for LiVSe$_2$. The plane-wave cutoff is set to $K_\mathrm{max} = 7.5 / R_{\mathrm{MT}}$. 

\section{Results and discussion}

\subsection{X-ray diffraction experiments at ambient pressure}

\begin{figure}
\includegraphics[width=86mm]{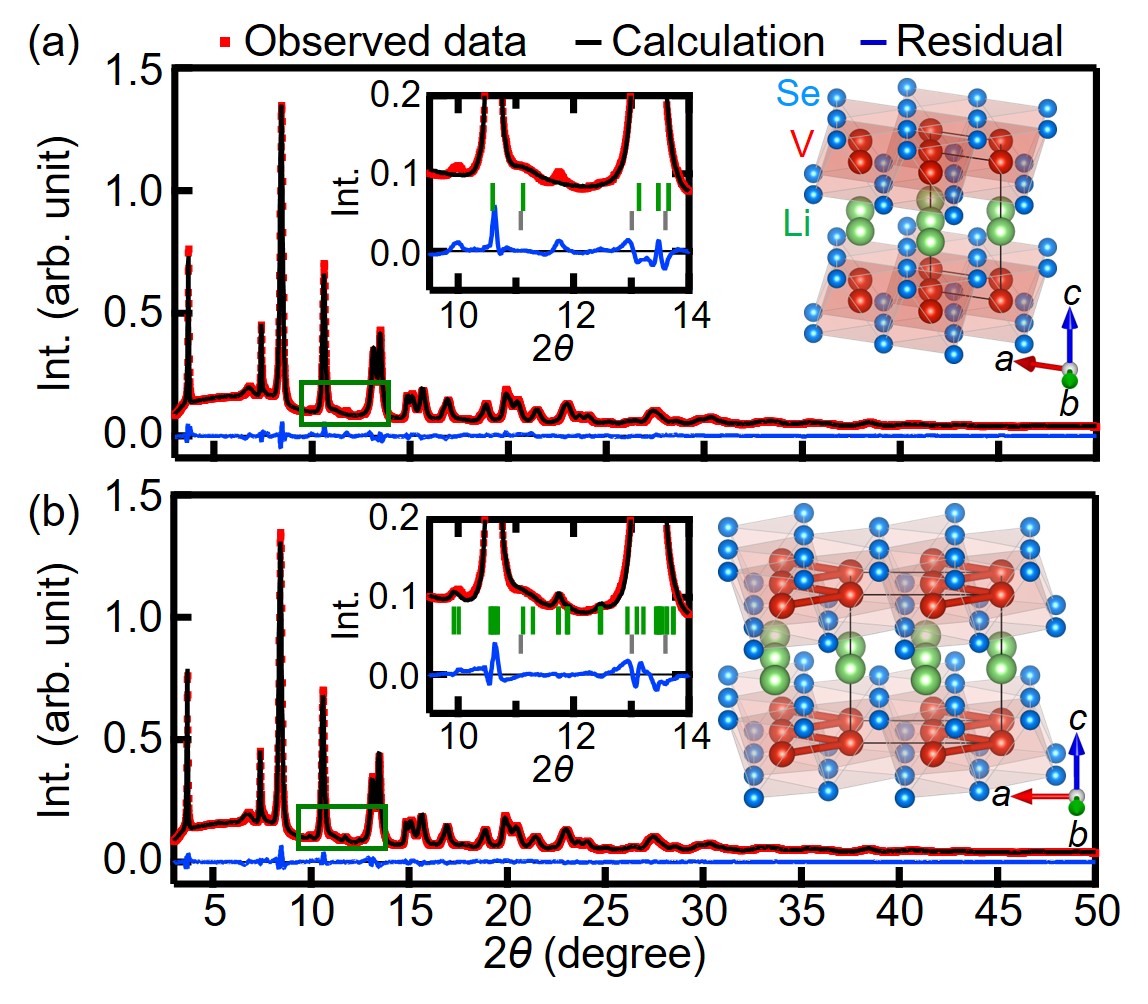}
\caption{\label{fig:rietveld} (a,b) Rietveld analysis at 100 K for assuming (a) a triangular lattice and (b) a zigzag chain structure. The insets show the area of the green box. The green ticks are the peak positions of the structure used for fitting, and the gray ticks are for impurity Li$_2$Se.}
\end{figure}

Figure~\ref{fig:rietveld}(a) shows the results of Rietveld analysis at 100 K performed assuming a triangular lattice structure of LiVSe$_2$ with space group $P\bar{3}m1$ and the presence of Li$_2$Se impurities. Although the simulation pattern seems to reproduce the diffraction data well, some peaks could not be indexed based on the structure model with a space group $P\bar{3}m1$. Assuming the structure of the monoclinic space group $Pm$ with vanadium forming zigzag chains, all peaks, including these unknowns, were fitted well, as shown in Fig.~\ref{fig:rietveld}(b). The obtained lattice parameters are $a_{\mathrm{m}}$ = 6.2407(6) $\mathrm{\AA}$, $b_{\mathrm{m}}$ = 3.5632(3) $\mathrm{\AA}$, $c_{\mathrm{m}}$ = 6.3425(2) $\mathrm{\AA}$, $\beta$ = 89.884(14)$^{\circ}$, $\alpha$ = $\gamma$ = 90$^{\circ}$. As can be seen from the fact that $a_{\mathrm{m}}/\sqrt{3}b_{\mathrm{m}}$ = 1.011 is close to 1 and $\beta$ is close to 90$^{\circ}$, the unit cell shape does not clearly show monoclinic distortion, which is the reason why diffraction pattern is fitted fairly well using the trigonal space group $P\bar{3}m1$. However, as indicated by the appearance of distinct superlattice peaks, the formation of the zigzag chains splits the V-V distance into three different types, as shown in the inset of Fig. \ref{fig:Phase_diag}.


\begin{figure}
\includegraphics[width=86mm]{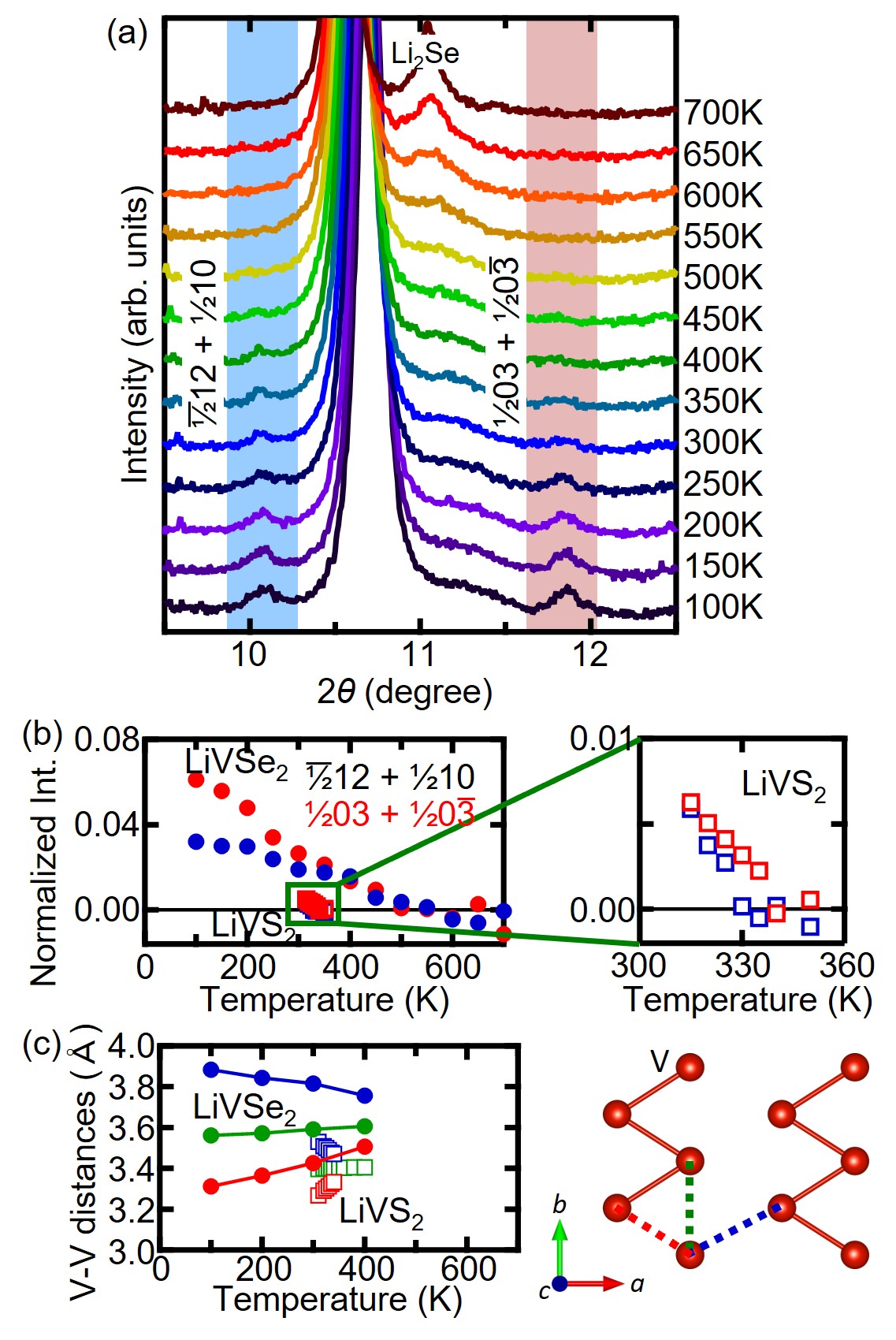}
\caption{\label{fig:XRD_Tdep} 
(a) Temperature dependence of the x-ray diffraction data of LiVSe$_2$ obtained at ambient pressure. The indices of each peak listed in the figure assume a triangular lattice structure. (b) Intensity ratios of the peaks $\bar{\frac{1}{2}}$12 + $\frac{1}{2}$10 (blue) and $\frac{1}{2}$03 + $\frac{1}{2}$0$\bar{3}$ (red) in LiVSe$_2$ and LiVS$_2$ (Ref. \cite{LiVS$_2$-2}) to the 001 fundamental peak. The values due to LiVS$_2$ are so small that they are enlarged and shown on the right. (c) Temperature dependence of the V-V distance obtained by Rietveld analysis. The colors plotted for each interatomic distance correspond to the dotted lines shown in the structure diagram to the right of the graph.}
\end{figure}

As shown in Fig.~\ref{fig:XRD_Tdep}(a), the intensity of the 
superlattice peaks that appear due to monoclinic distortion gradually decrease with increasing temperature and are no longer observed as a distinct peak above 500 K. For quantitative evaluation, the intensities of the peaks at $\bar{\frac{1}{2}}$12 + $\frac{1}{2}$10 and $\frac{1}{2}$03 + $\frac{1}{2}$0$\bar{3}$ normalized by the 001 peak are shown in Fig.~\ref{fig:XRD_Tdep}(b). The intensities of the superlattice peaks decay linearly with increasing temperature, reaching zero at around 500 K. Corresponding to the decay of these intensities, the splits of in-plane V-V interatomic distance shown in Fig.~\ref{fig:XRD_Tdep}(c) gradually decrease with increasing temperature.

The structural features of the zigzag chain state of LiVSe$_2$ shown above are similar to those of the high-temperature metallic phase of LiVS$_2$ \cite{LiVS$_2$-2}. In LiVS$_2$, even just above the metal-insulator transition temperature of 314 K, where the monoclinic distortion related to the zigzag chain is largest, $\beta$ is 89.956(3)$^{\circ}$ and $a_{\mathrm{m}}/\sqrt{3}b_{\mathrm{m}}$ retains to 1.001, which is the indicator of the magnitude of the monoclinic distortion. Thus, the monoclinic distortion feature is hardly apparent in the unit cell shape. Nevertheless, a large vanadium displacement of $\sim$0.13 $\mathrm{\AA}$ appears with zigzag chain formation.


The reason for these strange structural features appearing in LiVS$_2$ is related to the fact that the zigzag chain order is not long-range enough, and the correlation length remains finite. This is evident from the experimental results in Fig.~1(a) in the reference \cite{LiVS$_2$-2}, where the vanadium displacement estimated by the average structure is always smaller than the local vanadium displacement estimated by the pair distribution function analysis (PDF analysis). If the correlation length of the zigzag chain is long enough, the magnitude of the vanadium displacement obtained from the average structural analysis should be equal to that obtained from the local structural analysis. These features are also evident in the shape of the unit cell as estimated by the average structure. If such a large vanadium displacement occurred, the unit cell would naturally be more distorted from trigonal. However, if the monoclinic distortion associated with vanadium displacement is only localized and the monoclinic domains are oriented in various directions with short correlation lengths, the average structure analysis will appear as if a trigonal lattice is realized. In LiVS$_2$, such monoclinic distortions are developed with “mesoscopic” correlation lengths of a few hundred \AA~ or more, so to speak, which is why they are observed as such strange structural features.


\subsection{XAFS experiments on LiV$X_2$ ($X$ = O, S, Se) series}

\begin{figure}
\includegraphics[width=86mm]{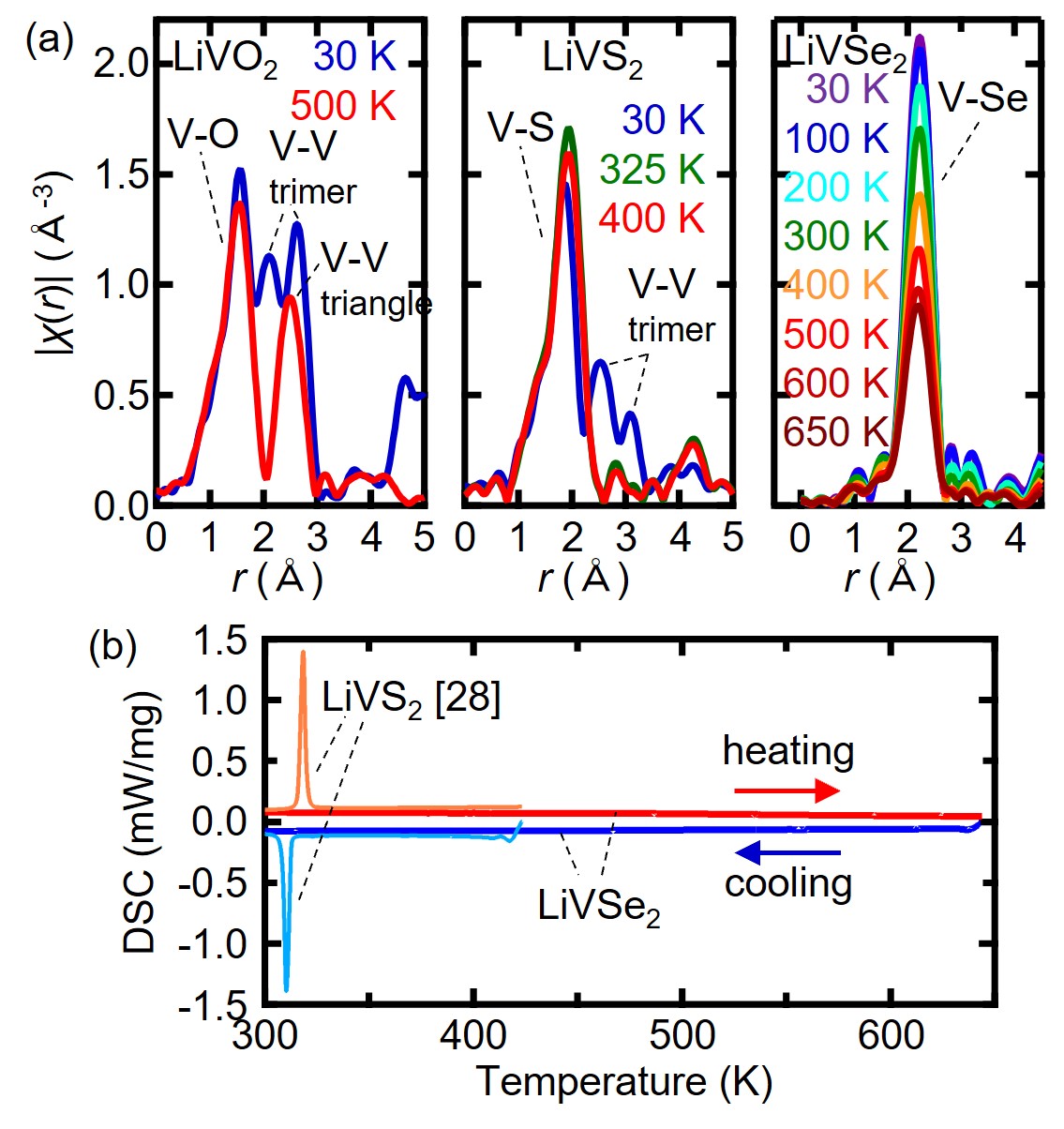}
\caption{\label{fig:XAFS} (a) The RDFs by XAFS measurements on LiV$X_2$. (b) Results of DSC measurements in LiVSe$_2$ and reference LiVS$_2$; results of DSC measurements in LiVS$_2$ are equal to those reported in previous studies \cite{LiVS$_2$-2}.}
\end{figure}

It is interesting that structural features similar to LiVS$_2$ were observed in LiVSe$_2$. This seems to suggest that the same features are realized in LiVSe$_2$ as in LiVS$_2$, where the correlation length of the zigzag chain remains finite and shortens with increasing temperature. To confirm this, we performed the vanadium $K$-edge XAFS measurements of three LiV$X_2$ compounds ($X$ = O, S, and Se) at various temperatures. Fig.~\ref{fig:XAFS}(a) shows the temperature dependencies of the radial distribution function (RDF) for three samples. This is a function of the probability of an atom at a distance $r$ from any given vanadium in the crystal structure, obtained by the Fourier transform of the XAFS spectrum. In the LiVS$_2$ data at 30 K, a peak corresponding to the V-S distance appears at $\sim$2 $\mathrm{\AA}$, and two peaks corresponding to the V-V distances associated with vanadium trimerization appear from 2.5 to 3 $\mathrm{\AA}$, as shown in Fig.~\ref{fig:Phase_diag}. It should be noted that the presence of two separate peaks has already been discussed in reference \cite{LiVS$_2$-1}. Strangely enough, once the temperature is raised to the high-temperature phase, the peaks corresponding to the V-V distance disappear completely, and only the peaks related to the V-S distance appear.

The absence of a peak corresponding to the V-V distance is consistent with the previous paper \cite{LiVS$_2$-1}, but the cause is not mentioned there. This may be since the correlation length of the zigzag chains of vanadium in the high-temperature phase LiVS$_2$ is much smaller than the beam diameter of 500 $\mu$m and shows fluctuating dynamics in time and space. These dynamics occur on the order of seconds, which is well faster than the measurement time of 60 s for XAFS spectra. Because of the relationship between the features exhibited by zigzag chains and the experimental conditions, the addition of spectra obtained from multiple zigzag chain structures with different orientations is observed in XAFS measurements. Then, due to the phase cancellation of the XAFS spectra, the peaks corresponding to the V-V distance are lost in the RDF. It should be noted that exactly the same RDF results are obtained at 325 K, where the displacement of vanadium from the zigzag chains is clearly observed in the average structure, and at 400 K, where no displacement is observed. This suggests that the local structure remains unchanged as the temperature increases and that the vanadium lattice similarly fluctuates at 325 and 400 K.

Similarly, in the spectrum of LiVSe$_2$, a peak corresponding to the V-Se distance is observed near $\sim$2.2 $\mathrm{\AA}$, but no peak corresponding to the V-V distance is observed at all. If the zigzag chain structure in LiVSe$_2$ forms a long-range order without fluctuating, three types of split V-V distances should be observed in RDF. Therefore, the present result indicates that the zigzag chain interactions are not sufficiently long-range ordered and disordered as in LiVS$_2$. In addition, interestingly, no clear change in the RDF appears well beyond around 500 K, where the superlattice peaks disappear in x-ray diffraction data. This clearly suggests that the zigzag chain order persists to high temperatures while becoming short-range ordered in LiVSe$_2$ as well.

It should be noted that the correlation length of the zigzag chain changes continuously with temperature, and therefore no clear signature is observed in the physical property measurements. As shown in Fig.~\ref{fig:XAFS}(b), DSC measurements of LiVSe$_2$ performed in the range of 300 - 650 K show no anomaly around 500 K. This is also the case in LiVS$_2$. In the DSC data of LiVS$_2$, the superlattice peaks disappear without any sign around 350 K, indicating that the cause of the disappearance of the superlattice peaks is similar in both cases.

The above experimental results might seem to indicate that the zigzag chains that appear as precursors to trimerization in LiVS$_2$ also persist in LiVSe$_2$, but this is not such a simple case. If the zigzag chain order is best developed near the trimer phase transition boundary, then the zigzag chain order of LiVS$_2$, which shows trimerization at low temperatures, should be more stable than that of LiVSe$_2$. However, as shown in Figs.~\ref{fig:XRD_Tdep}(b) and (c), the zigzag chain order in LiVSe$_2$ is stable at higher temperatures with larger vanadium displacements. It should be pointed out that in LiVO$_2$, which has the highest trimer transition temperature in the LiV$X_2$ series, no zigzag chain order is reported to appear at high temperatures \cite{LiVO2-kj}. In fact, as shown in Fig.~\ref{fig:XAFS}(a), XAFS measurement of LiVO$_2$ clearly shows a single V-V distance peak in the high-temperature phase as well as two V-V distances originating from the trimerization that appear in the low-temperature phase. The appearance of a peak corresponding to the V-V distance in the high-temperature phase is also reported in Ref. \cite{LiVO2_xafs}. This indicates that the vanadium lattice disorder that appears in LiVS$_2$ and LiVSe$_2$ does not occur in the high-temperature phase of LiVO$_2$. This indicates that this anomalous zigzag chain state appears universally and stably over a range of the metallic phase of LiVS$_2$ and LiVSe$_2$, not only in the vicinity of the trimer phase, as shown in Fig.~\ref{fig:Phase_diag}.


\subsection{Partial DOS calculations of LiVSe$_2$}

\begin{figure}
\includegraphics[width=86mm]{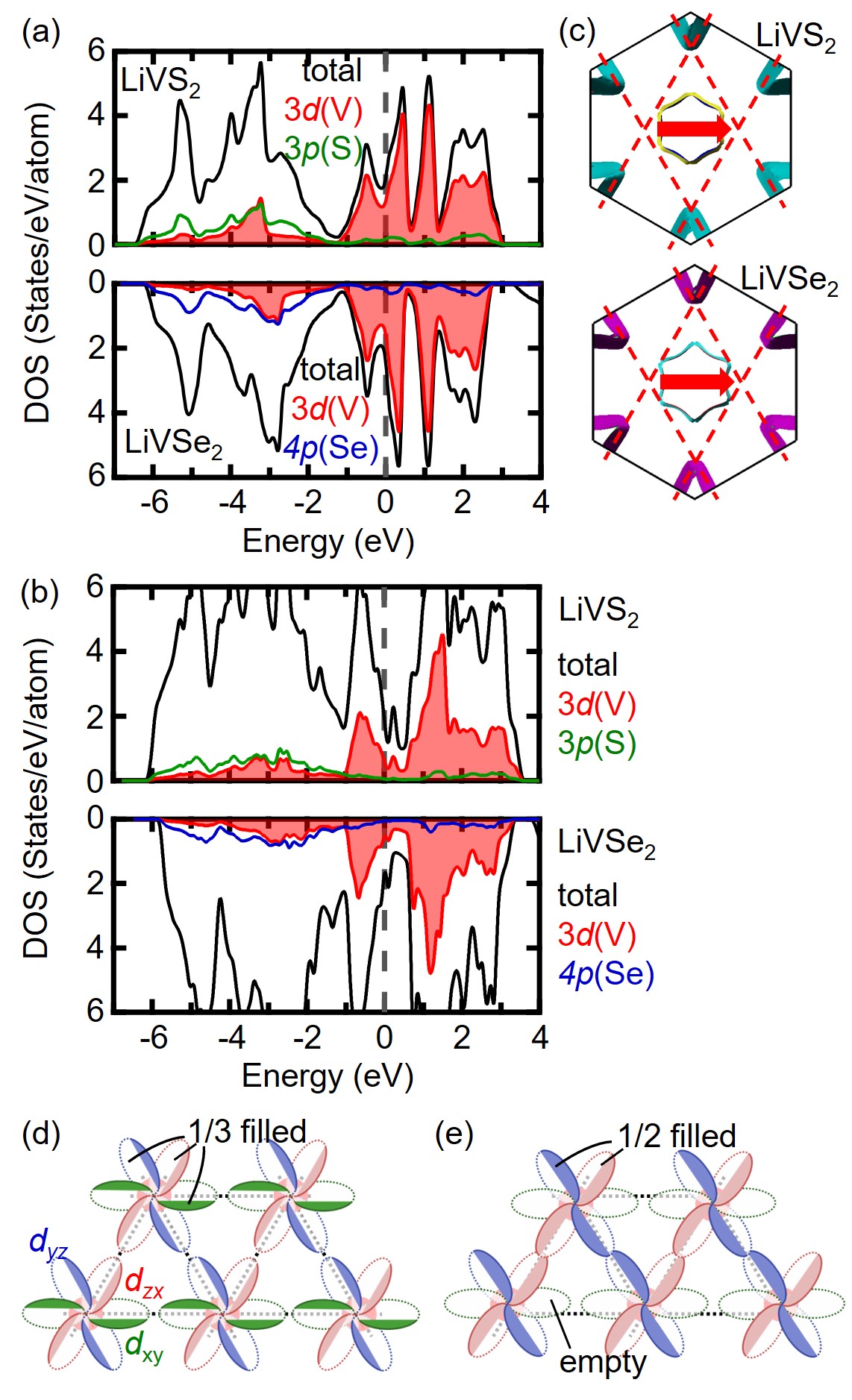}
\caption{\label{fig:DOS} (a) Partial-DOSs of LiVS$_2$ (upper) and LiVSe$_2$ (lower) assuming a triangular lattice structure. (b) Partial-DOSs of LiVS$_2$ (upper) and LiVSe$_2$ (lower) assuming a zigzag chain structure. (c) Shapes of Fermi surfaces for LiVS$_2$ (upper) and LiVSe$_2$ (lower) with a triangular lattice structure. The figures of LiVS$_2$ are equal to those shown in the previous study \cite{LiVS$_2$-2}. (d) Relationship between $d_{xy}$, $d_{yz}$, and $d_{zx}$ orbitals. Electrons occupy 1/3 of each orbital. (e) Zigzag chain formation due to the change of filling by charge transfer.}
\end{figure}

The above structural studies show that zigzag chain ordering is realized in both metallic phases of LiVS$_2$ and LiVSe$_2$, persistently surviving with the correlation length becoming shorter when the temperature increases. A unique feature of this system is that the zigzag chains are more stabilized at LiVSe$_2$, which is farther from the metal-insulator transition boundary than LiVS$_2$ in the phase space shown in Fig.~\ref{fig:Phase_diag}. Partial DOS calculations were performed to determine the cause of these characteristics. Figure~\ref{fig:DOS}(a) is the DOS computed on the basis of the space group $P\bar{3}m1$ realized as an average structure at high temperatures in LiVS$_2$ and LiVSe$_2$. The calculation result for LiVS$_2$ is essentially identical to that shown in \cite{LiVS$_2$-2}. LiVS$_2$ and LiVSe$_2$ have similar DOS shapes with a large population due to the 3$d$ orbitals of vanadium near the Fermi surface. However, due to the difference in lattice parameters associated with the difference between S(3$p$) and Se(4$p$), the energy width of the $d$ orbitals is slightly narrower in LiVSe$_2$. This will be important in later discussions.

On the other hand, Fig.~\ref{fig:DOS}(b), calculated for the space group $Pm$ structure with zigzag chain structure, shows a clear decrease in the density of states near the Fermi surface for both LiVS$_2$ and LiVSe$_2$. This happens to be due to the nesting of the Fermi surfaces that occurs as a result of zigzag chain formation. In the structure of the space group $P\bar{3}m1$ of LiVS$_2$, it has been discussed that there is an instability of the Fermi surface defined by the nesting vector $Q$ = $a^*$/2 as shown in Fig.~\ref{fig:DOS}(c), which causes zigzag chain formation corresponding to vanadium displacements. A similar instability of the Fermi surface is expected to exist in LiVSe$_2$, which may lead to the zigzag chain formation in LiVSe$_2$. However, the Fermi surfaces have not completely disappeared in both cases, which is consistent with the high-temperature phase of LiVS$_2$ and LiVSe$_2$ being essentially metallic.


The reason why the zigzag chain appears more stable in LiVSe$_2$, even though the partial DOS of LiVS$_2$ and LiVS$_2$ are very similar, may be due to the difference in the energy width of the $d$ orbital. In the trigonal structure, the $d_{xy}$, $d_{yz}$, and $d_{zx}$ orbitals that consist of $t_{2g}$ orbitals are triply degenerate, and the electron filling of these three orbitals are equal to 1/3, as shown in Fig.~\ref{fig:DOS}(d). If the triangular lattice is distorted to form a zigzag chain, in principle the electron filling must change from ($d_{xy}$,$d_{yz}$,$d_{zx}$)=(1/3,1/3,1/3) to a filling of ($d_{xy}$,$d_{yz}$,$d_{zx}$)=(0,1/2,1/2), as shown in Fig.~\ref{fig:DOS}(e). This has been pointed out by Rovira and Whangbo \cite{Rovira_Whangbo}, who argued that narrower $t_{2g}$ orbitals require less energy for electron transfer, thus stabilizing the zigzag chain state. LiVS$_2$ and LiVSe$_2$ actually maintain metallic conduction in the zigzag chain state. Therefore, ($d_{xy}$,$d_{yz}$,$d_{zx}$)=(0,1/2,1/2) is not strictly realized and some electrons are considered to remain in the $d_{xy}$ orbitals. However, the difference in bandwidth as pointed out by Rovira and Whangbo is found in our theoretical calculations, as shown in Fig.~\ref{fig:DOS}(a), which may be the reason why more stable zigzag chains appear in LiVSe$_2$.


\subsection{X-ray diffraction under high pressure}

\begin{figure*}
\includegraphics[width=175mm]{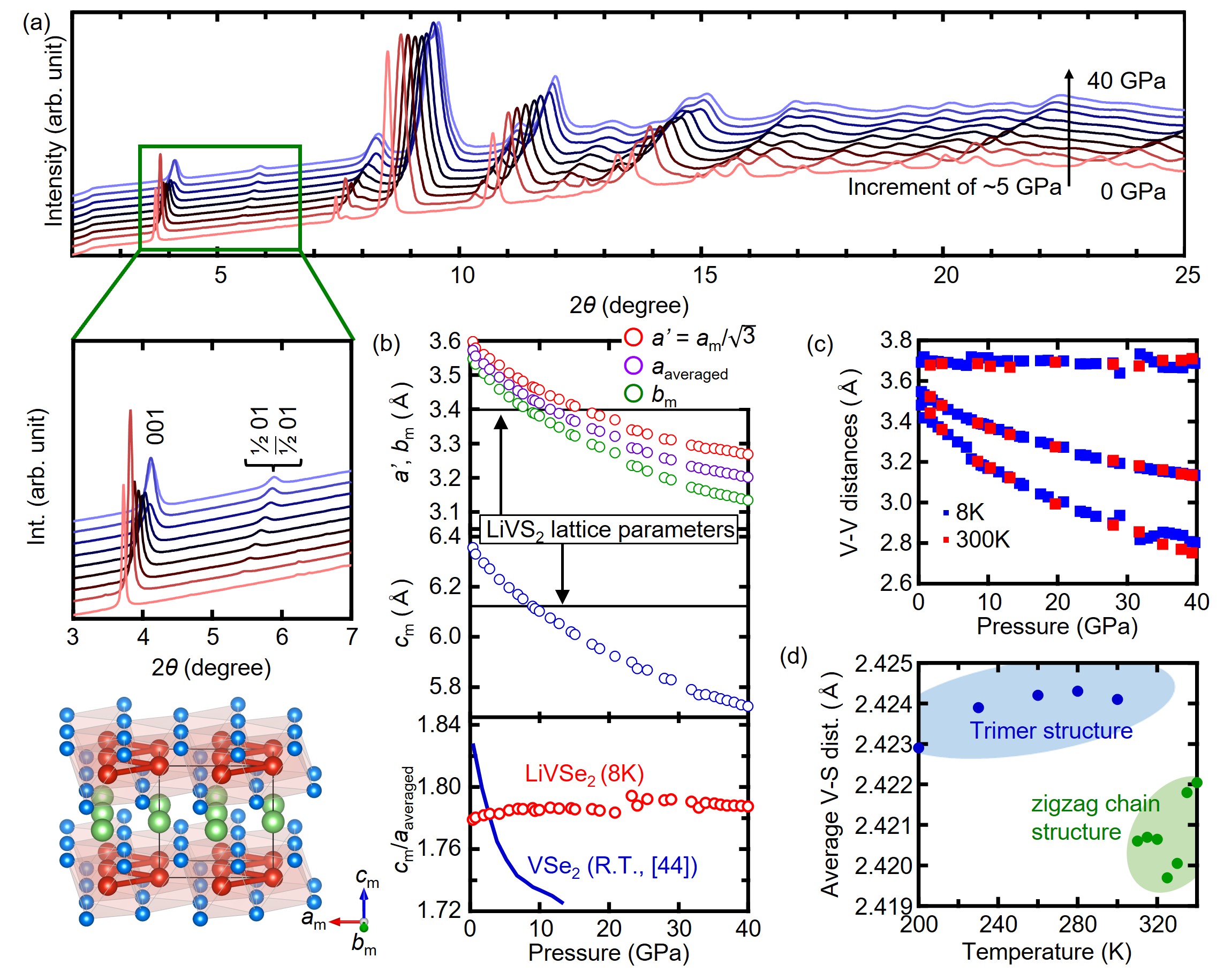}
\caption{\label{fig:pressure} (a) Pressure dependence of diffraction data. The figure shows diffraction data from ambient pressure to 40 GPa in increments of about 5 GPa. An enlarged view of the low-angle data is shown below. (b) Pressure dependencies of the lattice parameters and $c/a$ at 8 K. $c/a$ for VSe$_2$ is taken from R. Sereika $et~al.$ \cite{VSe$_2$_pressure} as a reference. (c) Pressure dependence of the estimated in-plane V-V distance. (d) Temperature dependence of averaged V-S distance in LiVS$_2$.}
\end{figure*}

Since the zigzag chain formation is associated with the instability of the Fermi surface, it is expected that applying pressure causes poor nesting of the Fermi surface. Instead, trimerization is stabilized by the effect of closer V-V distance and stronger $d$-$d$ hybridization. Based on these expectations, we performed diffraction experiments under pressure. Fig.~\ref{fig:pressure}(a) shows the diffraction data at several pressure points at 8 K. When pressure is applied, the diffraction peak clearly shifts to the high-angle side, indicating that the lattice parameter is decreased, as summarized in the Le Bail analysis results in Fig.~\ref{fig:pressure}(b). The pressure dependence of the diffraction data in Fig.~\ref{fig:pressure}(a) shows that the intensity of the superlattice peaks from the zigzag chain structure gradually increases with increasing pressure, suggesting that, contrary to expectations, the zigzag chain state stabilizes under pressure. The Rietveld analysis is necessary to estimate the amount of atomic displacement of V that causes zigzag chains, but it is difficult to perform accurate analyses because of the reduced crystallinity and higher background under pressure. Therefore, we used the following method to estimate the atomic positions in the crystal. First, Rietveld analysis was performed on the data at 20.8(1) GPa, where the peaks are relatively sharp, and the superlattice intensity is reasonable. Using the structural parameters obtained from this analysis and the ratio of fundamental and superlattice intensity, we estimated the structural parameters for each data. The Supplemental Material \cite{SI} describes the method's details. As shown in Fig.~\ref{fig:pressure}(c), the splitting of the V-V interatomic distance at 8 K increases continuously under pressure, reaching as much as 0.33 $\mathrm{\AA}$ (10.6\%) at 40.0(1) GPa. Furthermore, as shown in Fig.~\ref{fig:XRD_Tdep}(c), while the V-V interatomic distance shows a strong temperature dependence at ambient pressure, there is no difference in the magnitude of displacement between 8 K and 300 K under pressure, indicating that the zigzag chains are rapidly stabilized under pressure.

There are two possible reasons for the unexpected stabilization of the zigzag chains. First, the Li ions between the VSe$_2$ layers prevented the dimensionality from decreasing under pressure. This is evident from the pressure dependence of $c/a$, a measure of dimensionality, shown in the lower part of Fig.~\ref{fig:pressure}(b). Generally, the $c/a$ value decreases rapidly with applied pressure in layered compounds because the van der Waals gap between layers is easily compressed. In VSe$_2$, which is included for reference, the value of $c/a$ decreases with pressure \cite{VSe$_2$_pressure}, but in LiVSe$_2$, contrary to expectations, it does not change. This behavior may be due to the insertion of Li ions between the VSe$_2$ layers, which prevents the van der Waals gap from compressing. The presence of Li in LiVSe$_2$ not only sets the stage for the formation of zigzag chains by changing vanadium to the $d^2$ electronic state, but also prevents the zigzag chains from becoming unstable under pressure by losing the two-dimensionality of the electronic structure, which is important for Fermi surface nesting. Second, the V-S distance shrinks under pressure, and $d$-$p$ hybridization is strengthened. As shown in Fig.~\ref{fig:pressure}(d), in LiVS$_2$ where vanadium trimerization appears at low temperatures, the average V-S distance is clearly increased by trimerization. This would indicate that trimerization appears from strong $d$-$d$ hybridization between ionic V$^{3+}$($d^2$) and therefore does not favor strong $d$-$p$ hybridization. Under high pressure in LiVSe$_2$, both $d$-$d$ and $d$-$p$ hybridization are enhanced. Although these two effects should be competing, the effect of reduced V-ionicity due to increased $d$-$p$ hybridization is more dominant in LiVSe$_2$, resulting in the trimer state not appearing under high pressure in LiVSe$_2$. Instead, the zigzag chain state stabilizes under high pressure.

\subsection{Discussion}

From the above experiments, it seems likely that the zigzag chain states appearing in the LiV$X_2$ system are not simply trimer precursors, but are characterized as competing electronic phases governed by various parameters such as Fermi surface nesting, band width, $d$-$p$ and $d$-$d$ hybridization. On the other hand, however, the feature that the correlation length of the zigzag chain state remains finite up to high temperatures indicates that the zigzag chain state is very different from the usual electronic phase, rather similar to the local distortion features that appear in the high-temperature phase of many systems where molecular formation occurs at low temperatures \cite{MgTi2O4_3,Li2RuO3_2,RuP2,LiRh2O4-2,AlV2O4-2, MTe2, NaTiSi2O6_2}. There seem to be several cases of local distortion phenomena that appear at high temperatures: First, the case in which molecular formation patterns that appear at low temperatures appear as short-range order. This is the most common case and has been discussed in Li$_2$RuO$_3$ \cite{Li2RuO3_2}, AlV$_2$O$_4$ \cite{AlV2O4-2}, LiRh$_2$O$_4$ \cite{LiRh2O4-2}, RuP \cite{RuP2}, etc. The recently reported dimer fluctuation of 1$T$-$M$Te$_2$ ($M$ = V, Nb, Ta) may also be a variant of this case \cite{MTe2}. Second, the case in which distortion patterns different from those of the low-temperature phase appear as short-range order. A typical example is CuIr$_2$S$_4$ \cite{CuIr2S4_2}, where it is argued that the dimer of the low-temperature phase does not appear as a short-range order, but instead the tetragonal distortion appears as a short-range order. Third, the case of LiV$X_2$ ($X$ = S, Se), where a completely different pattern from the low-temperature phase appears with a finite correlation length. Among these, the LiV$X_2$ system is also unusual in that it forms an order with a long correlation length on the mesoscopic scale, and it undoubtedly provides a unique playground that sets it apart from other material systems that exhibit molecular formation. Through further exploration of materials forming short-range order, we expect that the physics of these molecule-forming systems will develop into an important research field related to various fields, including quantum liquid crystals and nonequilibrium physics.


\section{Summary}

In summary, we have performed structural studies and theoretical calculations of layered LiV$X_2$ and found that the zigzag chain state that appears in the high-temperature phase of LiVS$_2$ is more stabilized in LiVSe$_2$. This indicates that the zigzag chain phase appears more stable in the phase space of LiV$X_2$ ($X$ = O, S, Se) in regions away from the trimer phase boundary. The zigzag chain phases found in this system are significantly different from the local distortion features that appear in conventional molecular formation systems, clearly indicating that molecular formation systems with a long history, as typified by the Verwey transition of magnetite, are still unique playgrounds with a rich variety of fascinating physics.\\

\begin{acknowledgments}
We are very grateful to R. Isoda and Dr. R. Ishii for their experimental support. The work leading to these results has received funding from the Grant in Aid for Scientific Research (No. JP17K17793, No. JP20H02604, No. JP20H01849, No. JP21K18599, No. JP21J21236, No. JP22KJ1521, and No. JP23H04104) and Research Foundation for the Electrotechnology of Chubu. This work was carried out under the Visiting Researcher’s Program of the Institute for Solid State Physics, the University of Tokyo, and the Collaborative Research Projects of Laboratory for Materials and Structures, Institute of Innovative Research, Tokyo Institute of Technology. Powder x-ray diffraction experiments were conducted at the BL02B2, BL10XU, and BL44B2 of SPring-8, Hyogo, Japan (Proposals No. 2020A1063, No. 2021A1111, No. 2021A1117, No. 2021B1136 and No. 2022A1167), and at the BL5S2 of Aichi Synchrotron Radiation Center, Aichi Science and Technology Foundation, Aichi, Japan (Proposals No. 202105170, No. 202106066, No. 2021L1002, and No. 2020L6002). XAFS experiments were conducted at the BL5S1 and BL11S2 of Aichi Synchrotron Radiation Center (Proposals No. 201901017, 201902049, 202201031, and 202302018).
\end{acknowledgments}

\appendix

\nocite{*}

\bibliography{references}

\end{document}